\documentclass[conference,a4paper]{IEEEtran}

\usepackage{amsmath, amsthm, amssymb}
\usepackage{amsfonts}
\usepackage{amsthm}
\usepackage{tikz}
\usetikzlibrary{shapes,arrows}
\usepackage{verbatim}
\usepackage[]{subfigure}
\theoremstyle{remark}

\usepackage{graphicx,relsize,verbatim,enumerate,xcolor}
\usepackage{pgfplots}

\usetikzlibrary{arrows,positioning,fit,backgrounds,shapes}
\usetikzlibrary{calc}

\newtheorem{lemma}{Lemma}
\newtheorem{theorem}{Theorem}
\newtheorem{definition}{Definition}

\newcommand{\distnarg}[2][p]{\ensuremath{\MakeUppercase{#1}_{#2}}}

\newcommand{\rightratearrow}[5]{
\draw ([yshift=#3] #1) to node [midway,above] {#5} ([yshift=#3,xshift=-#4] #2) -- ++(0,#3) -- (#2);
\draw ([yshift=-#3] #1) -- ([yshift=-#3,xshift=-#4] #2) -- ++(0,-#3) -- (#2);
                            }

\tikzstyle{arw}=[->,>=latex]
\tikzstyle{node}=[rectangle,draw,outer sep=0pt,minimum width=1.7cm, minimum height=8mm]

\newcommand{\E}{{\mathbb{E}}}
\newcommand{\br}[1]{{\left(#1\right)}}
\newcommand{\setbr}[1]{{\left\{#1\right\}}}
\newcommand{\abs}[1]{{\left|#1\right|}}

\newcommand{\R}{{\mathbb{R}}}

\newcommand{\defeq}{{~\triangleq~}}

\newcommand{\inv}{{^{-1}}}

\newcommand{\X}{{\mathcal{X}}}
\newcommand{\Y}{{\mathcal{Y}}}
\newcommand{\Z}{{\mathcal{Z}}}
\newcommand{\U}{{\mathcal{U}}}
\newcommand{\V}{{\mathcal{V}}}
\newcommand{\C}{{\mathcal{C}}}
\newcommand{\B}{{\mathcal{B}}}
\newcommand{\s}{{\mathcal{S}}}

\newcommand{\J}{{\mathcal{J}}}
\newcommand{\supp}{{\mbox{support}}}
\newcommand{\mtiny}[1]{{\mbox{\tiny{$#1$}}}}

\newcommand{\ex}{{\exists}}
\newcommand{\e}{{\epsilon}}
\newcommand{\es}{{\emptyset}}

\newcommand{\setb}[1]{{\left\{#1\right\}}}

\newcommand{\normtv}[1]{{\left\|#1\right\|_{\mbox{\tiny{$TV$}}}}}

\begin{document}

\sloppy

%% Paper Title
%% You can use linebreaks \\ within to get better formatting as
%% desired. 
\title{Secure Cascade Channel Synthesis} 

\author{
  \IEEEauthorblockN{Sanket Satpathy and Paul Cuff}
  \IEEEauthorblockA{Dept. of Electrical Engineering\\
    Princeton University\\
    Princeton, USA\\
    Email: \{satpathy,cuff\}@princeton.edu}
}

%% Create the title:
\maketitle

%% Abstract: 
%% For the final version of the accepted paper, please make sure you
%% remove the comment "THIS PAPER IS ELIGIBLE FOR THE STUDENT PAPER
%% AWARD."
%%
\begin{abstract}
  We investigate channel synthesis in a cascade setting where nature provides an iid sequence $X^n$ at node 1. Node 1 can send a message at rate $R_1$ to node 2 and node 2 can send a message at rate $R_2$ to node 3. Additionally, all 3 nodes share bits of common randomness at rate $R_0$. We want to generate sequences $Y^n$ and $Z^n$ along nodes in the cascade such that $(X^n,Y^n,Z^n)$ appears to be appropriately correlated and iid even to an eavesdropper who is cognizant of the messages being sent. We characterize the optimal tradeoff between the amount of common randomness used and the required rates of communication. We also solve the problem for arbitrarily long cascades and provide an inner bound for cascade channel synthesis without an eavesdropper.
\end{abstract}

\section{Introduction}

Since Shannon introduced the notion of mutual information, many attempts have been made to measure the correlation between random variables\cite{Gacs,Wyner,coord,Cuff1}. Each of them is relevant in the context of the operational questions they address. New life has been breathed into these attempts by a line of inquiry into measures of correlation with a view to perform synthesis under a total variation constraint, known as strong coordination\cite{coord} or channel synthesis\cite{DCS}. The optimal tradeoff between communication and common randomness was derived by Cuff\cite{Cuff1} and Bennett et al.\cite{reverse2} for the case of two random variables and the results have been independently rediscovered and extended in other work\cite{Gohari1,Gohari2,Gohari3,Gohari4,Winter,DCS}. One particularly pleasing aspect of the above tradeoff was that it recovered two familiar measures of correlation as the required rate of communication - mutual information and Wyner's common information\cite{Wyner} in the presence of abundant and no common randomness, respectively.

Requiring that the synthesized joint distribution be close to the desired joint distribution in total variation is a more stringent constraint than empirical coordination\cite{Haim,coord} i.e.\ jointly typical input and output sequences. On the other hand, we enjoy the benefit of the synthesized sequences being immune to statistical tests designed to detect iid correlated sequences\cite{DCS}. This aspect of the results coupled with the nice properties of total variation as a metric such as a bound on entropy\cite[Theorem 17.3.3]{Cover} has led to applications in secrecy\cite{Cuff3,Cuff4,Winter05} and game theory\cite{Cuff1,Cuff2,DCS}. We refer the interested reader to \cite{DCS,Bloch2} for a more thorough discussion of the properties of total variation and comparison with other metrics.

\tikzstyle{block} = [draw, fill=blue!20, rectangle, 
    minimum height=2em, minimum width=4em]
\tikzstyle{sum} = [draw, fill=blue!20, circle, node distance=1cm]
\tikzstyle{input} = [coordinate]
\tikzstyle{output} = [coordinate]
\tikzstyle{pinstyle} = [pin edge={to-,thin,black}]

\begin{figure}[ht]
\begin{center}
\resizebox{3in}{1.1in}{% The block diagram code is probably more verbose than necessary
\begin{tikzpicture}[scale=1,auto, node distance=1cm,>=latex']
    % We start by placing the blocks
    \node [input, name=input] {};
    \node [block, below of=input, node distance=1.5cm] (node1) {$F_n$};
    \node [block, right of=node1,
            node distance=4cm] (node2) {$G_n$};
%    \node [block, right of=node2, pin={[pinstyle]90:Disturbances},
%            node distance=3cm] (node3) {$H_n$};
    \node [block, right of=node2,
            node distance=4cm] (node3) {$H_n$};
    % We draw an edge between the node2 and node3 block to 
    % calculate the coordinate u. We need it to place the measurement block. 
    \draw [->,double] (node2) -- node[name=r2] {$nR_2$ bits} (node3);
    \draw [] (node3) -- node[name=j2] {$J_2$} (node2);
    \node [output, below of=node2, node distance=1.5cm] (output1) {};
    \node [output, below of=node3, node distance=1.5cm] (output2) {};
    \node [draw=none, above of=node2, node distance = 2cm] (commrand) {$nR_0$ bits};

    % Once the nodes are placed, connecting them is easy. 
    \draw [draw,<-] (node1) -- node {$X^n$} (input);
    \draw [->,double] (node1) -- node {$nR_1$ bits} (node2);
    \draw [] (node2) -- node {$J_1$} (node1);
    \draw [->] (node2) -- node [name=yn] {$Y^n$}(output1);
    \draw [->] (node3) -- node [name=zn] {$Z^n$}(output2);
    \draw [->,double] (commrand) -- (node1);
    \draw [->,double] (commrand) -- (node2);
    \draw [->,double] (commrand) -- (node3);
%    \draw [->] (y) |- (commrand);
%    \draw [->] (commrand) -| node[pos=0.99] {$-$} 
%        node [near end] {$y_m$} (node1);
\end{tikzpicture}}
\caption{The iid sequence $X^n$ is given by nature. Messages $J_1$ and $J_2$ are sent along the cascade at rates $R_1$ and $R_2$. Common randomness at rate $R_0$ is shared by all 3 nodes. We want $(X^n,Y^n,Z^n)$ iid correlated and independent of the messages $(J_1,J_2)$.}\label{sccs}
\end{center}
\end{figure}
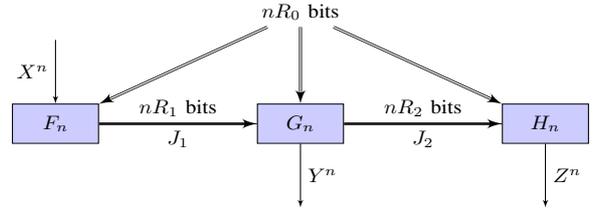

This particular work is inspired by the suite of extensions presented in \cite{DCS} with a focus on secrecy applications. To extend these applications to general networks, it is desirable to have a theory of strong coordination over arbitrary networks\cite{Gohari1,Gohari4}. The cascade channel synthesis problem shown in Fig.\ref{sccs} provides a starting point for such a theory. Distributed networks for control and sensing are seemingly ubiquitous - the power grid, road networks, server farms and the internet are a few prominent examples\cite{control}. In many control settings, one would like the actions at various nodes to be independent of the communication so the actions cannot be anticipated by malicious eavesdroppers. We consider the problem of synthesizing sequences that appear to be iid and appropriately correlated even to an onlooker who can see messages sent over the communication channel, but does not have access to the source of common randomness.

The cascade structure for communication is especially relevant for large distributed networks where there is a cost per unit distance associated with sending messages. In such a setting, it would be more economical for nodes to forward appropriate messages locally in a cascade fashion rather than having a central node talk to all the other nodes. Also in some settings, there might be a hierarchy among the nodes that inherits the cascade structure (for example, in parallel computation interaction takes place in a master-slave hierarchy). In Section IV, we characterize the optimal rate region for the task assignment problem, where 3 out of $m\ge3$ tasks have to be distributed uniformly at random to the 3 nodes, with the first node's tasks chosen at random by nature.

In this work, we assume that every node in the cascade has sufficient local randomness. We provide a precise description of the problem in Section II and present a characterization of the optimal rate region in Section III. One unexpected feature of the optimal region is that there is no loss of generality in assuming that the second message is a function of the first i.e.\ the local randomness used in synthesizing $Y^n$ is not essential to correlating $Z^n$ with $(X^n,Y^n)$. However, it is precisely this feature that proves to be our undoing for the corresponding scheme in the cascade channel synthesis problem with no eavesdropper as discussed in Section VII.

\section{Preliminaries and Problem Definition}

\subsection{Notation}

We represent both random variables and probability distribution functions with capital letters, but only letters $P$ and $Q$ are used for the latter. If it is not clear from context, we include subscripts to denote that $P_{Y|X}(y|x)$ is the conditional distribution of the random variable $Y$ given the random variable $X$. We may abbreviate this as $P(y|x)$ or $P_{Y|X}$. Also, we use the script letter $\X\ni x$ to denote the alphabet of $X$. Sequences of random variables $X_1,\ldots,X_n$ are denoted by $X^n$. The set $\setb{1,\ldots,m}$ is denoted by $[m]$.

Markov chains are denoted by $X-Y-Z$ implying the factorization $P_{XYZ}=P_{XY}P_{Z|Y}$. We define the total variation distance as
\begin{equation}
\normtv{P_X-Q_X}\defeq\frac{1}{2}\sum_x\abs{P(x)-Q(x)}.\label{tv}
\end{equation}\vspace{-.4cm}

\subsection{Problem-Specific Definitions}

We have the iid source $X^n\sim\prod_{i=1}^nQ_X$ and we would like to \emph{synthesize} the channel $Q_{YZ|X}$. Messages sent along the cascade communication links are denoted by $J_1\in[2^{nR_1}]$ and $J_2\in[2^{nR_2}]$. The common randomness shared by all nodes $K$ is uniformly distributed on $[2^{nR_0}]$ and independent of $X^n$.

\begin{definition}
A $(2^{nR_0},2^{nR_1},2^{nR_2},n)$ \emph{secure cascade channel synthesis (SCCS) code} consists of randomized encoding functions\vspace{-0.3cm}
\begin{IEEEeqnarray*}{rCl}
F_n&:&\X^n\times[2^{nR_0}]\to[2^{nR_1}],\\
G_n^{(enc)}&:&[2^{nR_1}]\times[2^{nR_0}]\to[2^{nR_2}],\vspace{-0.1cm}
\end{IEEEeqnarray*}
and randomized decoding functions
\begin{IEEEeqnarray*}{rCl}
G_n^{(dec)}&:&[2^{nR_1}]\times[2^{nR_0}]\to\Y^n,\\
H_n&:&[2^{nR_2}]\times[2^{nR_0}]\to\Z^n.
\end{IEEEeqnarray*}
We have $J_1=F_n(X^n,K),J_2=G_n^{(enc)}(J_1,K)$, $Y^n=G_n^{(dec)}(J_1,K)$ and $Z^n=H_n(J_2,K)$.
\end{definition}

Note that node 2 has both encoding and decoding capability. The \emph{induced joint distribution} of a $(2^{nR_0},2^{nR_1},2^{nR_2},n)$ SCCS code is the joint distribution $P_{X^n,Y^n,Z^n,K,J_1,J_2}$ as per the above specifications.

\begin{definition}
A sequence of $(2^{nR_0},2^{nR_1},2^{nR_2},n)$ SCCS codes for $n\ge1$ is said to \emph{achieve} $Q_{YZ|X}$ if the induced joint distributions have marginals that satisfy
\begin{equation}
\lim_{n\to\infty}\normtv{P_{X^nY^nZ^nJ_1J_2}-P_{J_1J_2}\prod_{t=1}^nQ(x_t,y_t,z_t)}=0.
\end{equation}
\end{definition}

\begin{definition}
A rate triple $(R_0,R_1,R_2)$ is said to be \emph{achievable} if there exists a $(2^{nR_0},2^{nR_1},2^{nR_2},n)$ SCCS code that achieves $Q_{YZ|X}$.
\end{definition}

\begin{definition}
The \emph{secure synthesis rate region} $\mathcal{C}$ is the closure of achievable triples $(R_0,R_1,R_2)$.
\end{definition}

\section{Main Result}
The characterization of the set of achievable rate triples is given in terms of the following set
\begin{equation}\s_{D}\defeq \setbr{\begin{array}{r c l}
\hspace{-.1cm}(R_0,R_1,R_2)\in\R^3&:&\ex P_{X,Y,Z,U,V}\in D \mbox{ s.t.}\\
R_1&\ge&I(X;U,V)\\
R_2&\ge&I(X;V)\\
R_0&\ge&I(X,Y,Z;U,V)
\end{array}},\label{result}\end{equation}where
\begin{equation}D\defeq\setbr{\begin{array}{r c l}
P_{X,Y,Z,U,V}&:&(X,Y,Z)\sim Q_XQ_{YZ|X},\\
&&X-(U,V)-Y,\\
&&(X,Y,U)-V-Z,\\
&&\abs{\V}\le\abs{\X}\abs{\Y}\abs{\Z}+3,\\
&&\abs{\U}\le\abs{\X}\abs{\Y}\abs{\Z}\abs{\V}+2
\end{array}}.\label{D}\end{equation}Also, let $D'=D\cap\setbr{P_{X,Y,Z,U,V}:H(V|U)=0}$ denote the restriction of $D$ to joint distributions where $V$ is a function of $U$.

\begin{theorem}
\begin{equation}\C=\s_D=\s_{D'}.\end{equation}\label{thm}\end{theorem}
\vspace{-0.6cm}

\section{Observations and Examples}

Note that the communication rate region of our result coincides with the rate region for empirical coordination in the cascade channel\cite{coord} since there must exist a realization of the shared randomness that yields good empirical coordination codes, in agreement with Theorem 2 of \cite{coord}. Also this observation consolidates the intuition that the onus of secrecy that we have taken on is borne primarily by the available common randomness.

The two node channel synthesis problem\cite{Cuff1,DCS} recovered mutual information and Wyner's common information as the required communication rates at the extremes of abundant and no common randomness, respectively. The problem we consider is that of secure coordination. We do not have constraints on the sum of communication and common randomness rates since we cannot afford to have public messages give away too much about our coordination scheme. The extremes of our region are not determined by the amount of common randomness available, but we see that the minimum rates follow the same trend as above.

By the data-processing inequality\cite{Cover}, the choice $(U,V)=(Y,Z)$ simultaneously minimizes both the communication rates at $R_1\ge I(X;Y,Z)$ and $R_2\ge I(X;Z)$. Also, the minimum achievable rate of common randomness is
\begin{equation}C_c(X;Y;Z)=\min_{(U,V):\substack{X-(U,V)-Y,\\(X,Y,U)-V-Z}}I(X,Y,Z;U,V),\label{casc}
\end{equation}
which can be viewed as a generalization of Wyner's common information in the cascade setting. Another straightforward generalization of Wyner's common information that has been considered in the literature\cite{comm} is\vspace{-0.1cm}
\begin{equation}C(X;Y;Z)=\min I(X,Y,Z;U),\label{wyn}
\end{equation}
where the minimum is over random variables $U$ such that given $U$, $X,Y$ and $Z$ are independent of each other. In fact, the two quantities are the same.
This is because the minimizers of \eqref{wyn} and \eqref{casc} are compatible with each other's Markov structures. For example, if $\hat{U}$ attains the minimum for \eqref{wyn}, then $(U,V)=(\es,\hat{U})$ satisfies the Markov chains in \eqref{casc}. Note that the communication and common randomness rates cannot be simultaneously minimized in general.\vspace{-0.1cm}

\subsection{Task Assignment}

We now compute our region for a toy problem of task assignment, where 3 out of $m\ge3$ tasks are to be assigned to the 3 nodes uniformly at random. To be precise, we consider a channel $Q_{YZ|X}$ that acts on $X$ uniformly distributed on $[m]$ i.e. $Q_X=m\inv$ and produces a pair $Y\neq Z$ uniformly distributed over all distinct pairs in $[m]\setminus\setbr{X}$. This is a generalization of the scatter channel example in \cite{DCS}.

As per \eqref{result}, we consider joint distributions $P_{X,Y,Z,U,V}\in D$. The Markov chains ensure that for each pair $(u,v)$ in the support of $(U,V)$, the conditional distributions $P_{X,Y|U=u,V=v}$ and $P_{X,Y,Z|V=v}$ factor as $P_{X|U=u,V=v}P_{Y|U=u,V=v}$ and $P_{X,Y|V=v}P_{Z|V=v}$ respectively. Also, these distributions have the constraint that the supports of $X,Y$ and $Z$ cannot intersect.

The above constraint dictates that $\abs{\supp(X,Y)}=a\in[m-1]\setminus\setbr{1}$ and that $\abs{\supp(Y)}=b\in[a-1]$, enumerating all possible sparsity patterns. Since we seek $U$ and $V$ to enforce the above Markov chains, consider $V$ to specify $\supp(X,Y)$ and $U$ to specify $\supp(Y)$. For each of the above categories, we have trivial bounds on conditional entropy:\vspace{-0.2cm}
\begin{IEEEeqnarray}{rCl}
H(X|V=v)&\le&\log a\\
H(X|U=u,V=v)&\le&\log(a-b)\\
H(X,Y,Z|U=v,V=v)&\le&\log(a-b)b(m-a).
\end{IEEEeqnarray}
The above inequalities give lower bounds on the required rates that are achieved by letting $U$ and $V$ be uniformly distributed over all appropriate supports of sizes $b$ and $a$ respectively, and selecting uniform distributions over supports. Thus, the rate region is given by the convex hull of the set\vspace{-0.3cm}

\begin{equation}\setbr{\begin{array}{r c l}
(R_0,R_1,R_2)\in\R^3&:&\ex a\in[m-1], b\in[a-1]\mbox{ s.t.}\\
R_1&\ge&\log\br{\frac{m}{a-b}}\\
R_2&\ge&\log\br{\frac{m}{a}}\\
R_0&\ge&\log\br{\frac{m(m-1)(m-2)}{(a-b)b(m-a)}}\\
\end{array}}.\end{equation}

The communication rates are minimized when $a=m-1$ and $b=1$ i.e.\ given $(U,V)$, the uncertainty is concentrated on $X$. On the other hand, the common randomness requirement is minimized when $b\approx\frac{m}{3}$ and $a\approx\frac{2m}{3}$ up to the nearest integer i.e.\ given $(U,V)$, the uncertainty is shared equally by $X,Y$ and $Z$. The tradeoff between information content of the messages and rate of common randomness is evident here.\vspace{-.1cm}

\section{Sketch of Converse}

Let $(R_0,R_1,R_2)$ be achievable. Then for $\e\in(0,1/4)$ there exists a $(2^{nR_0},2^{nR_1},2^{nR_2},n)$ secure channel synthesis code with an induced joint distribution $P_{X^n,Y^n,Z^n,K,J_1,J_2}$ such that
\begin{equation}
\lim_{n\to\infty}\normtv{P_{X^nY^nZ^nJ_1J_2}-P_{J_1J_2}\prod_{l=1}^nQ(x_l,y_l,z_l)}<\e.
\end{equation}
We shall use the random variable $T$ uniformly distributed on $[n]$, as a time index. First, we use the triangle inequality and \cite[Lemma V.1]{DCS} to note that
$$\normtv{P_{X^nY^nZ^nJ_1J_2}-P_{J_1J_2}P_{X^nY^nZ^n}}\le\cdots$$
\begin{equation}2\normtv{P_{X^nY^nZ^nJ_1J_2}-P_{J_1J_2}\prod_{l=1}^nQ(x_l,y_l,z_l)}<2\e.\vspace{-.1cm}\label{indep}\end{equation}

\subsection{Entropy Bounds}

We will need the following bounds on entropy in terms of total variation\cite[Lemma VI.3]{DCS}. If the joint distribution of $(X^n,Y^n,Z^n)$ is close in total variation to an iid distribution as assumed, then we have
\begin{align}
\sum_{t=1}^nI_P(X_t,Y_t,Z_t;X^{t-1},Y^{t-1},Z^{t-1})&\le& ng(\e),\label{tv1}\\
I_P(X_T,Y_T,Z_T;T)&\le&ng(\e),\label{tv2}
\end{align}
where
\begin{equation}
g(\e)\defeq 4\e\log\br{\frac{\abs{\X}\abs{\Y}\abs{\Z}}{\e}}.
\end{equation}
Note that $\lim_{\e\downarrow0}g(\e)=0$.

\subsection{Approximate Rate Region}

We use standard information-theoretic inequalities, the physical constraint $X^n-(J_1,K)-J_2$ and the fact that $X^n$ and $K$ are independent to bound the communication rates:\vspace{-.2cm}
\begin{IEEEeqnarray}{rClCl}
nR_1&\ge&H(J_1)&\ge& H(J_1|K)\\
&\ge& I(X^n;J_1|K)&=&I(X^n;J_1,J_2,K),\\
nR_2&\ge&H(J_2)&\ge& H(J_2|K)\\
&\ge& I(X^n;J_2|K)&=&I(X^n;J_2,K).\vspace{-.2cm}
\end{IEEEeqnarray}
Finally, we bound $R_0$:\vspace{-.2cm}
\begin{IEEEeqnarray}{rCl}
nR_0&\ge&H(K)\ge H(K|J_1,J_2)\\
&\ge& I(X^n,Y^n,Z^n;K|J_1,J_2)\\
&\ge&I(X^n,Y^n,Z^n;J_1,J_2,K)-g(2\e)\\
&\ge&\sum_{t=1}^nI(X_t,Y_t,Z_t;J_1,J_2,K)-g(2\e)-ng(\e)\\
&\ge&nI(X_T,Y_T,Z_T;J_1,J_2,K|T)-(n+1)g(\e)\\
&\ge&nI(X_T,Y_T,Z_T;J_1,J_2,K,T)-(2n+1)g(\e).\label{corr}
\end{IEEEeqnarray}
Theorem 17.3.3 of \cite{Cover} coupled with \eqref{indep}  was used in the third line and the other steps follow from \eqref{tv1} and \eqref{tv2}. The single-letterization for the communication rates is done similarly. Making associations $U'=J_1$ and $V'=(J_2,K,T)$, we see that the rates and Markov chains in \eqref{result} and \eqref{D} are satisfied up to the correction in \eqref{corr}.

Using a generalized Carath\'{e}odory theorem, we can show existence of a distribution $\Gamma_{XYZUV}$ with $(U,V)$ satisfying the cardinality bounds and Markov chains as in \eqref{D}, but using \cite[Lemma VI.2]{DCS} we only have that\vspace{-.1cm}
\begin{equation}\normtv{\Gamma_{XYZ}-Q_XQ_{YZ|X}}<\e,\vspace{-.1cm}\end{equation}
yielding approximate versions of $\s_D$ and $D$. The final step is to take $\e\downarrow0$ and use a compactness argument as in \cite[Lemma VI.5]{DCS} to conclude that the above region coincides with \eqref{result} in the limit.\vspace{-.2cm}

\subsection{Comment on Converse}

Note that the above arguments go through with the associations $U=(J_1,J_2,K,T)$ and $V=(J_2,K,T)$, establishing that there is no loss of generality in assuming that $V$ is a function of $U$ in the rate region description. In the argument for the cardinality bound, we find it helpful to rewrite $(X,Y,U)-V-Z$ as the equivalent constraint $I(X,Y,U;Z|V)=0$. This lets us first bound $\abs{\V}$ without worrying about $\abs{\U}$.

\section{Sketch of Achievability}

\subsection{Cloud-Mixing}

Given a channel $ Q_{V|U}$, we want to synthesize $V^n\sim\prod Q_V$ at the output. However, we would like to do it using a small codebook of $U^n\sim\prod Q_U$ codewords. How large does the codebook need to be? The cloud-mixing lemma provides the answer. It is implied by results on resolvability\cite{resolvability} and goes back to Wyner's work\cite{Wyner}. The key ingredients of our proof are the following lemmas.

\begin{lemma}[Lemma IV.1 in \cite{DCS}]
For any discrete distribution $ Q_{UV}$ and $n\ge1$, let $\B^{(n)}=\setbr{U^n(m)}_{m=1}^{2^{nR}}$ be a codebook of sequences each drawn independently from $\prod_{k=1}^n Q_U(u_k)$. For a fixed codebook, define\vspace{-0.2cm}
\begin{equation}
P(v^n)=2^{-nR}\sum_{m=1}^{2^{nR}}\prod_{k=1}^n Q_{V|U}(v_k|U_k(m)).\vspace{-0.2cm}
\end{equation}
Then if $R>I(V;U)$,\vspace{-0.1cm}
\begin{equation}
\lim_{n\to\infty}\E\normtv{P(v^n)-\prod_{k=1}^n Q_V(v_k)}=0,\vspace{-0.1cm}
\end{equation}
where the expectation is with respect to the random codebook $\B^{(n)}$.\label{cloud1}
\end{lemma}

\begin{lemma}[Lemma IV.17 in \cite{DCS}]
For any discrete distribution $ Q_{UVZ}$ and $n\ge1$, let $\B_1^{(n)}=\setbr{U^n(l)}_{l=1}^{2^{nR_1}}$ and $\B_2^{(n)}=\setbr{V^n(m,l)}_{m=1}^{2^{nR_2}}$ for $l\in[2^{nR_1}]$ be codebooks of sequences drawn from $\prod_{i=1}^nQ_U$ and $\prod_{i=1}^nQ_{V|U_i(l)}$ respectively. For fixed codebooks, define\vspace{-0.1cm}
\begin{equation}
P(z^n)=2^{-n(R_1+R_2)}\sum_{l=1}^{2^{nR_1}}\sum_{m=1}^{2^{nR_2}}\prod_{k=1}^n Q_{Z|UV}(z_k|U_k(l),V_k(m,l)).\vspace{-0.1cm}
\end{equation}
If we have 
\begin{IEEEeqnarray}{rCl}
R_1&>&I(U;Z),
\end{IEEEeqnarray}
\begin{IEEEeqnarray}{rCl}
R_2&>&I(U,V;Z)-H(U),\\
R_1+R_2&>&I(U,V;Z),
\end{IEEEeqnarray}
then
\begin{equation}
\lim_{n\to\infty}\E\normtv{P(z^n)-\prod_{k=1}^n Q_Z(z_k)}=0,
\end{equation}
where the expectation is with respect to the random codebooks.\label{cloud2}
\end{lemma}\vspace{-.5cm}

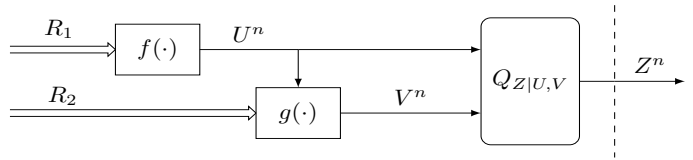
\begin{figure}[ht]
\begin{center}
\resizebox{3.5in}{0.8in}{\begin{tikzpicture}[node distance=2cm]
 \node (src1)   [coordinate] {};
 \node (src2)   [coordinate,below=1cm of src1] {};
 \node (chv)    [coordinate,below=5mm of src1] {};
 \node (label1) [coordinate,right=2cm of src1] {};
 \node (enc1)   [node,minimum width=12mm,right=15mm of src1] {$f(\cdot)$};
 \node (enc2)   [node,minimum width=12mm,right=35mm of src2] {$g(\cdot)$};
 \node (split1) [coordinate] at (src1 -| enc2.center) {};
 \node (chh)    [coordinate,right=27mm of enc2] {\distnarg[ Q]{V|U}};
 \node (ch)     [node,minimum width=14mm,minimum height=2cm,rounded corners] at (chv -| chh) {\distnarg[ Q]{Z|U,V}};
 \node (dashline) [coordinate,right=5mm of ch] {};
 \node (dash1)  [coordinate] at (src1 -| dashline) {};
 \node (dash2)  [coordinate] at (src2 -| dashline) {};
 \node (outline)  [coordinate,right=1cm of dashline] {};
 \node (out1)   [coordinate] at (src1 -| outline) {};
 \node (out2)   [coordinate] at (src2 -| outline) {};

 \rightratearrow{src1}{enc1.west}{1.5pt}{3pt}{$R_1$}{3pt};
 \draw[draw=none] (src2) to node [midway,above] {$R_2$} (src2 -| enc1.west);
 \rightratearrow{src2}{enc2.west}{1.5pt}{3pt}{}{3pt};
 \draw (enc1) to node [midway,above] {$U^n$} (split1);
 \draw[arw] (split1) to (src1 -| ch.west);
 \draw[arw] (split1) to (enc2);
 \draw[arw] (enc2) to node [midway,above] {$V^n$} (src2 -| ch.west);
 \draw (ch) to (ch.center -| dashline);
 \draw[arw] (chv -| dashline) to node [midway,above] {$Z^n$} (chv -| out2);
 \draw[dashed] ([yshift=7mm] dash1) to ([yshift=-7mm] dash2);
\end{tikzpicture}}
\caption{{\em Two-encoder Example:}  Sufficient rates for producing an i.i.d. $Z^n$ sequence in this two encoder system are given in Lemma~\ref{cloud2}.}
\label{figure cloud extension two encoders}
\end{center}
\end{figure}\vspace{-.8cm}

\subsection{Existence of Achievability Codes}

Assume that $(R_0,R_1,R_2)$ is in the interior of $\s_D$. Then there exists a distribution $ Q_{XYZUV}\in D$ such that the rates in \eqref{result} are satisfied. For $n\ge1$, let $(K,J_1,J_2)$ be uniformly distributed on $[2^{nR_0}]\times[2^{n(R_1-R_2)}]\times[2^{nR_2}]$. Application of Lemma \ref{cloud1} with $(U,V)$ in place of $U$ and Lemma \ref{cloud2} with $U$, $V$ interchanged and $(X,R_2,R_1-R_2)$ in place of $(Z,R_1,R_2)$ ensures the existence of codebooks $\B^{(n)}=\setbr{U^n(k,j_1,j_2),V^n(k,j_2)}_{(k,j_1,j_2)\in\mathcal{K}\times\J_1\times\J_2}$ such that
\begin{IEEEeqnarray}{rCl}
\lim_{n\to\infty}\normtv{P(x^n,y^n,z^n)-\prod_{t=1}^nQ(x_t,y_t,z_t)}&=&0,
\label{synth}
\end{IEEEeqnarray}
\begin{IEEEeqnarray}{rCl}
\lim_{n\to\infty}\normtv{P(x^n,k)-2^{-nR_0}\prod_{t=1}^nQ_{X}(x_t)}&=&0
\end{IEEEeqnarray}
respectively, where $P(x^n,y^n,z^n)$ and $P(x^n,k)$ are marginals derived from the joint distribution
$$
P(x^n,y^n,z^n,k,j_1,j_2)=2^{-n(R_0+R_1)}\times\cdots$$ 
$$\br{\prod_{t=1}^n Q(x_t,y_t|U_t(k,j_1,j_2),V_t(k,j_2)) Q(z_t|V_t(k,j_2))}.$$
Note that the Markov chains in place allow the separation of syntheses of $Z^n$ from $(X^n,Y^n,U^n)$, and of $X^n$ from $Y^n$. The messages that determine the $U^n,V^n$ codewords are compatible with the physical constraints imposed by the cascade structure. By satisfying all the individual demands of the problem, we have indirectly shown the existence of channel synthesis codes. Finally, note that fixing $(j_1,j_2)\in\J_1\times\J_2$ still leaves us with a sufficient rate of common randomness $R_0>I(X,Y,Z;U,V)$ for \eqref{synth} to hold by Lemma \ref{cloud1}. Thus, there must exist good codebooks for secure coordination such that\vspace{-0.1cm}
\begin{equation}
\lim_{n\to\infty}\normtv{P_{X^nY^nZ^nJ_1J_2}-2^{-nR_1}\prod_{t=1}^nQ(x_t,y_t,z_t)}=0.\nonumber
\end{equation}
Finally, note that given a secure strong coordination code, redefining $(U,V)$ to be $((U,V),V)$ we see that the corresponding distribution in $D'$ yields the same rates.\vspace{-0.2cm}

\subsection{Comment on Achievability}

Our scheme requires local randomization at all nodes - please refer to \cite{DCS} for a quantitative treatment of local randomness in channel synthesis. Also, observe that while it is intuitive to think of the common randomness as a one-time pad on the messages, we do not need to use such a construction in our proof. On the other hand, it seems desirable to have a more direct achievability scheme for channel synthesis with explicit constructions. Some attempts have been made in this direction\cite{Gohari5,Bloch1}.\vspace{-.1cm}

\section{Extensions}

\subsection{Arbitrarily Long Cascades}

Our main result of Theorem \ref{thm} can be readily extended to secure channel synthesis of a distribution $Q_{Y_1,\ldots,Y_{m-1}|X}$ for a cascade with $m\ge3$ nodes, with communication rates $R_1,\ldots,R_{m-1}$ on the links of the cascade and common randomness shared by all nodes at rate $R_0$.

Let $C_k^m\defeq(C_k,\ldots,C_m)$ for $C\in\setb{R,U,Y}$, $1\le i\le m-1$ and $1\le j\le m-2$ below. The optimal rate region is
\begin{equation}\s_{D_\mtiny{m}}\defeq \setbr{\begin{array}{r c l}
R_0^{m-1}\in\R^m&:&\ex P_{X,Y_1^{m-1},U_1^{m-1}}\in D_{\mbox{\tiny{$m$}}} \mbox{ s.t.}\\
R_i&\ge&I(X;U_i^{m-1})\\
R_0&\ge&I(X,Y_1^{m-1};U_1^{m-1})
\end{array}},\label{result2}\end{equation}where
\begin{equation}D_{\mbox{\tiny{$m$}}}\defeq\setbr{\begin{array}{r c l}
P_{X,Y_1^{m-1},U_1^{m-1}}&:&(X,Y_1^{m-1})\sim Q_XQ_{Y_1^{m-1}|X},\\
&&X-U_1^{m-1}-Y_1,\\
&&(X,Y_1^j,U_1^j)-U_{j+1}^{m-1}-Y_{j+1},\\
&&H(U_{i}^{m-1}|U_i)=0,
\end{array}}\end{equation}
with the cardinality bounds
\begin{equation}
\abs{\U_i}\le\abs{\X}\br{\prod_{k=1}^{m-1}\abs{\Y_k}}\br{\prod_{k=i+1}^{m-1}\abs{\U_k}}+m+i-2.
\label{card}\end{equation}
The proof follows the same steps as the main result, using an appropriate generalization of Lemma \ref{cloud2} to derive the communication rates - namely an application of Lemma \ref{cloud1} followed by repeated applications of \cite[Corollary IV.8]{DCS}.\vspace{-0.1cm}

\subsection{No Eavesdropper}

The cascade  channel synthesis problem with no eavesdropper does not ask that $(X^n,Y^n,Z^n)$ and $(J_1,J_2)$ are independent, but only that
\begin{equation}
\lim_{n\to\infty}\normtv{P_{X^nY^nZ^n}-\prod_{t=1}^nQ(x_t,y_t,z_t)}=0.\label{reg}
\end{equation}
Note that a good secure channel synthesis code is also a good channel synthesis code\cite[Lemma V.1]{DCS}. For the channel synthesis problem with two nodes, a simple modification of rates for the secure synthesis region is optimal for channel synthesis - namely that the rate constraint on common randomness can be shared by the communication rate as well. In the same spirit, we can replace the constraint on $R_0$ in \eqref{result} with the constraints\vspace{-.4cm}
\begin{IEEEeqnarray}{rCl}
R_1+R_0&\ge&I(X,Y,Z;U,V)\label{con1}\\
R_2+R_0&\ge&I(X,Y,Z;V),\label{con2}\vspace{-.2cm}
\label{noeav}\end{IEEEeqnarray}
which can be shown to be achievable, but not optimal. Consider the case when $X$ is independent of $(Y,Z)$. An application of Lemma \ref{cloud1} implies that the constraint on $(R_2+R_0)$ alone is sufficient. So let us select  $U=\es$ and $V$ independent of $X$ such that $Y-V-Z$. All rate constraints except \eqref{con1} and \eqref{con2} vanish so that using $R_2>0$ will force us to use $R_1>0$ at the optimal rates, since $V$ is a function of $U$. This region is not optimal. We suspect that another random variable $W$ such that $(X,Y,U)-(V,W)-Z$ and $U-V-W$ might be required in order to convey information about the local randomness used to synthesize $Y^n$.\vspace{-.2cm}

%% Appendix:
%% If needed a single appendix is created by
%\appendix
%% If several appendices are needed, then the command
%\appendices
%% in combination with further \section-commands can be used.

%% Use \section* for acknowledgement
\section*{Acknowledgment}

The authors would like to thank Curt Schieler for helpful discussions. This work is supported by the National Science Foundation (grant CCF-1116013) and the Air Force Office of Scientific Research (grant FA9550-12-1-0196).\vspace{-.1cm}

%% References:
%% We recommend the usage of BibTeX:
%%
%\bibliographystyle{IEEEtran}
%\bibliography{definitions,bibliofile}
%%
%% where we here have assume the existence of the files
%% definitions.bib and bibliofile.bib.
%% BibTeX documentation can be obtained at:
%% http://www.ctan.org/tex-archive/biblio/bibtex/contrib/doc/
%%
%%
%%
%% Or manual references (pay attention to consistency!):
%\begin{thebibliography}{1}
%\bibitem{shannon1948}
%  C.~E. Shannon, ``A mathematical theory of communication,''
%  \emph{Bell System Techn. J.}, vol.~27, pp. 379--423 and 623--656,
%  Jul. and Oct. 1948. 
%\end{thebibliography}

\bibliographystyle{ieeetr}

\bibliography{sccs}

\begin{thebibliography}{10}

\bibitem{Gacs}
P.~G\'{a}cs and J.~Korner, ``{Common information is far less than mutual
  information},'' {\em Probl. Inform. Control}, vol.~2, no.~2, pp.~149--162,
  1973.

\bibitem{Wyner}
A.~Wyner, ``The common information of two dependent random variables,'' {\em
  IEEE Trans. Inf. Theor.}, vol.~21, pp.~163--179, Sept. 2006.

\bibitem{coord}
P.~Cuff, H.~H. Permuter, and T.~M. Cover, ``Coordination capacity,'' {\em IEEE
  Trans. on Info. Theory}, vol.~56, no.~9, pp.~4181--4206, 2010.

\bibitem{Cuff1}
P.~Cuff, ``Communication requirements for generating correlated random
  variables,'' in {\em ISIT}, pp.~1393--1397, 2008.

\bibitem{DCS}
P.~Cuff, ``Distributed channel synthesis,'' {\em CoRR}, vol.~abs/1208.4415,
  2012.

\bibitem{reverse2}
C.~H. Bennett, I.~Devetak, A.~W. Harrow, P.~W. Shor, and A.~Winter, ``Quantum
  reverse shannon theorem,'' Tech. Rep. arXiv:0912.5537, Dec 2009.

\bibitem{Gohari1}
A.~Gohari and V.~Anantharam, ``Generating dependent random variables over
  networks,'' in {\em ITW}, pp.~698 --702, Oct. 2011.

\bibitem{Gohari2}
A.~Gohari, M.~H. Yassaee, and M.~R. Aref, ``Secure channel simulation,'' {\em
  CoRR}, vol.~abs/1207.3513, 2012.

\bibitem{Gohari3}
M.~H. Yassaee, A.~Gohari, and M.~R. Aref, ``Channel simulation via interactive
  communications,'' in {\em ISIT}, pp.~3053--3057, 2012.

\bibitem{Gohari4}
F.~Haddadpour, M.~H. Yassaee, A.~Gohari, and M.~R. Aref, ``Coordination via a
  relay,'' in {\em ISIT}, pp.~3048--3052, 2012.

\bibitem{Winter}
A.~{Winter}, ``{Compression of sources of probability distributions and density
  operators},'' {\em eprint arXiv:quant-ph/0208131}, Aug. 2002.

\bibitem{Haim}
T.~M. Cover and H.~H. Permuter, ``Capacity of coordinated actions,'' in {\em
  ISIT}, pp.~2701 --2705, June 2007.

\bibitem{Cover}
T.~M. Cover and J.~A. Thomas, {\em Elements of information theory (2. ed.)}.
\newblock Wiley, 2006.

\bibitem{Cuff3}
P.~Cuff, ``A framework for partial secrecy,'' in {\em GLOBECOM}, pp.~1--5,
  2010.

\bibitem{Cuff4}
P.~Cuff, ``Using a secret key to foil an eavesdropper,'' in {\em Allerton},
  pp.~1405 --1411, 29 2010-oct. 1 2010.

\bibitem{Winter05}
A.~Winter, ``Secret, public and quantum correlation cost of triples of random
  variables,'' in {\em ISIT 2005}, pp.~2270--2274, 2005.

\bibitem{Cuff2}
P.~Cuff, ``State information in bayesian games,'' in {\em Allerton}, 2009.

\bibitem{Bloch2}
M.~Bloch and J.~N. Laneman, ``Secrecy from resolvability,'' {\em CoRR},
  vol.~abs/1105.5419, 2011.

\bibitem{control}
R.~Murray, K.~Astrom, S.~Boyd, R.~Brockett, and G.~Stein, ``Future directions
  in control in an information-rich world,'' {\em Control Systems, IEEE},
  vol.~23, pp.~20 -- 33, apr 2003.

\bibitem{comm}
W.~Liu, G.~Xu, and B.~Chen, ``The common info. of n dependent random
  variables,'' {\em CoRR}, vol.~abs/1010.3613, 2010.

\bibitem{resolvability}
T.~Han and S.~Verdu, ``Approximation theory of output statistics,'' {\em Info.
  Theory, IEEE Trans. on}, vol.~39, pp.~752 --772, may 1993.

\bibitem{Gohari5}
M.~H. Yassaee, M.~R. Aref, and A.~Gohari, ``Achievability proof via output
  statistics of random binning,'' {\em CoRR}, vol.~abs/1203.0730, 2012.

\bibitem{Bloch1}
M.~R. Bloch, L.~Luzzi, and J.~Kliewer, ``Strong coordination with polar
  codes,'' {\em CoRR}, vol.~abs/1210.2159, 2012.

\end{thebibliography}

\end{document}